\documentclass[conference]{IEEEtran}

\IEEEoverridecommandlockouts

\usepackage{cite}

\ifCLASSINFOpdf
   \usepackage[pdftex]{graphicx}
\else
   \usepackage[dvips]{graphicx}
   \DeclareGraphicsExtensions{.eps}
\fi

\usepackage{grffile}
\usepackage{epstopdf}

\usepackage{subfig}

\usepackage[cmex10]{amsmath}
\usepackage{color}
\usepackage{array}
\usepackage{hyperref}
\begin{document}

\title{CRDSA, CRDSA++ and IRSA: Stability and Performance Evaluation}

\author{\IEEEauthorblockN{Alessio Meloni
\thanks{\copyright 2012 IEEE. The IEEE copyright notice applies.\ DOI: 10.1109/ASMS-SPSC.2012.6333080
}and Maurizio Murroni}
\IEEEauthorblockA{DIEE - Department of Electrical and Electronic Engineering\\
University of Cagliari\\
Piazza D'Armi, 09123 Cagliari, Italy\\
Email: \{alessio.meloni\}\{murroni\}@diee.unica.it}
}

\maketitle

\begin{abstract}
In the recent past, new enhancements based on the well established Aloha technique (CRDSA, CRDSA++, IRSA) have demonstrated the capability to reach higher throughput than traditional SA,  in bursty traffic conditions and without any need of coordination among terminals. In this paper, retransmissions and related stability for these new techniques are discussed. A model is also formulated in order to provide a basis for the analysis of the stability and the performance both for finite and infinite users population. This model can be used as a framework for the design of such a communication system.
\end{abstract}

\IEEEpeerreviewmaketitle

\section{Introduction}
Since its birth more than 40 years ago, ALOHA \cite{AbramsonALOHA1} \cite{AbramsonALOHA2} has established as one of the most well-known protocols for packet-switching wireless networks. Since then, several papers have been published on this random access technique. Some papers are concerned with the issue of improving the throughput (e.g. providing slots for synchronized transmission from users \cite{RobertsALOHA} or providing diversity by sending the same packet more than once \cite{DiversityALOHA}) while other papers concentrate on the study of the stability \cite{STAB1} \cite{STAB2} and on the retransmission policies to ensure that the communication is stable \cite{STABcontrol}. In fact, when retransmissions are considered, a feedback is created between the transmitter and the receiver. Therefore, depending on the arrival rate of packets to be transmitted and on the Packet Loss Ratio, the possibility of an overloaded or saturated channel is present. For this reason, studies on stability of such a communication scenario and on possible policies to have a channel working in optimal conditions have been of big interest.
Recently, a new technique named Contention Resolution Diversity Slotted Aloha (CRDSA) has been introduced \cite{CRDSA1}. This technique allows to reach values of throughput up to $0.55 \frac{pkt}{slot}$ using an approach similar to Diversity Slotted Aloha \cite{DiversityALOHA} (i.e. every packet is sent twice in different slots) and adding a Successive Interference Cancellation (SIC) process at the receiver in order to attempt restoring collided packets (details of how this is done will be illustrated in Section II). Afterwards, the same concept has been extended to more than two instances\footnote{The term \textit{instances} is used towards the paper with the meaning of "total number $l$ of transmissions for the same packet in different slots".} per packet (CRDSA++) \cite{CRDSA2} and also to irregular number of packet repetitions (IRSA) \cite{IRSA1}.
As a consequence of these new enhancements, also the study of the stability of these new techniques is of big interest. Therefore the objective of the study shown in this paper is the development of a model able to evaluate the stability for these new techniques when retransmissions are considered. To do so, we refer to \cite{STAB2} in which a performance evaluation of Slotted Aloha was presented. In particular, the mentioned paper analyzes system stability behaviour through the development of analytic models based on the so called \textit{equilibrium contour}, that represents those points for which the expected number of successful packets is equal to the average number of newly generated packets, so that the overall communication is stable (i.e. the expected number of packets sent is the same at any time).
In the followings, we adapt this model to our case to provide a useful tool for the study of the stability in CRDSA, CRDSA++ and IRSA. Moreover, a framework for the optimization will be introduced. The starting point and the aim of this paper are similar to those in \cite{KisslingStab}. However, while the graphical representation used by Kissling represents the drift of the communication as the one in \cite{STAB1}, our representation is based on the one in \cite{STAB2}. The details for this different representation choice will be clear throughout the paper, but the main point is basically the possibility to graphically distinguish a curve influenced by the retransmission probability and a load line determined by the number of users and their probability of transmission so that these dependencies are separated and it is easier to understand how the communication behaviour changes when changing those values. Moreover, in \cite{KisslingStab} the goal of showing agreement between simulations and expected analytical drift is reached by averaging simulations results over a consistent number of trials for any possible initial state (in terms of number of backlogged users). In this paper instead, the result of single simulations with various settings are presented, aiming at showing the outcomes in terms of backlogged users, throughput and so on, frame after frame and starting from the initial state with no backlogged users. This allows to empirically understand how and how much the communication values move around the expected behaviour. Finally, an average packet delay analysis is also introduced.

\section{System Overview and Problem Statement}
\begin{figure}[t!]
\includegraphics [width=0.95 \columnwidth] {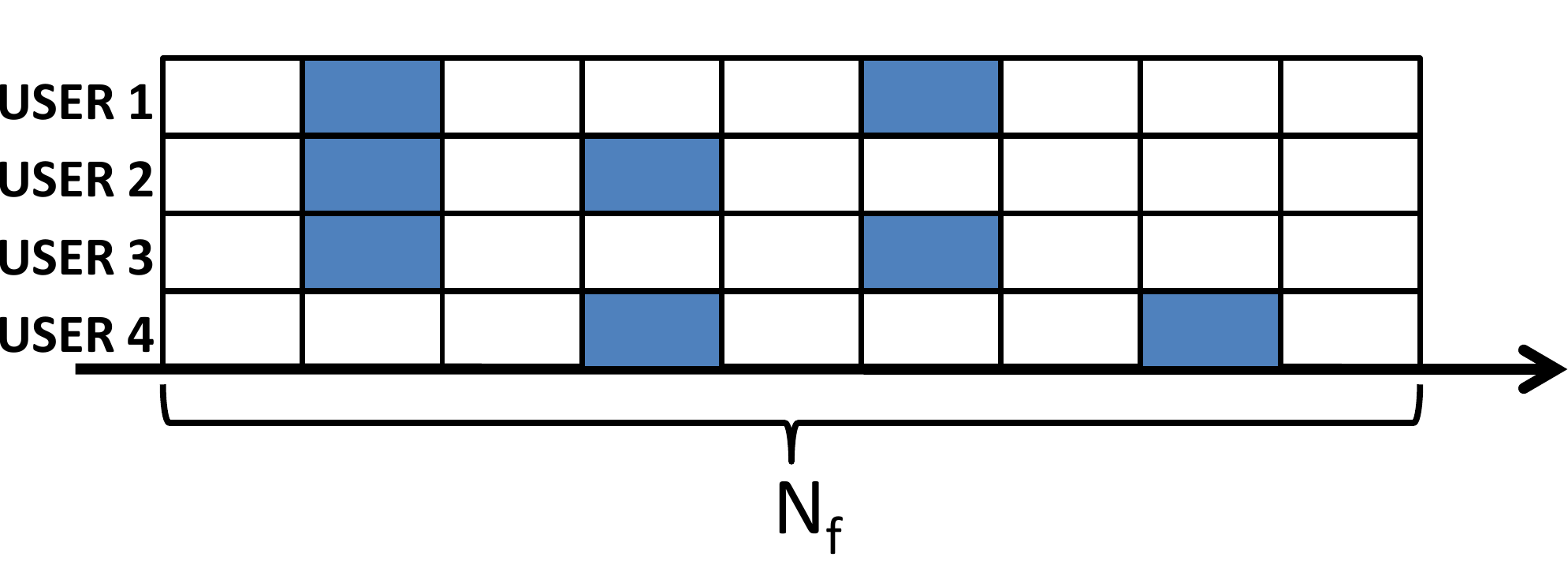}
\caption{\small{Example of frame at the receiver for CRDSA (2 instances per packet). Plain slots indicate that a transmission occurred for that user in that slot.}}
\label{Fig1}
\end{figure}

Consider a multi-access channel populated by a total number of users M (finite or infinite). Users are synchronized so that the channel is divided into slots and $N_f$ consecutive slots are grouped to constitute a so called frame. The probability that, at the beginning of a frame, an idle user has a packet to transmit is $p_0$. When a frame starts, users having a packet to transmit place $l$ instances of their packet over the $N_f$ slots of the frame. The number of instances $l$ can be either the same for each packet (regular \textit{burst degree distribution}\footnote{The \textit{burst degree distribution} is defined as the probability distribution to have a certain number of instances for a certain packet.}) \cite{CRDSA1} \cite{CRDSA2} or not (irregular \textit{burst degree distribution}) \cite{IRSA1}. Packet instances are nothing else than redundant copies as in \cite{DiversityALOHA} except for the fact that each instance contains a pointer to the location of the other instances. These pointers are used in order to attempt restoring collided packets at the receiver by means of Successive Interference Cancellation (SIC). Consider Figure~\ref{Fig1}, representing a frame at the receiver for the case of 2 instances per packet (CRDSA). Throughout the paper, we assume perfect interference cancellation and channel estimation, which means that the only cause of disturbance for the correct reception of packet's instances is interference among them. Moreover, FEC and possible power unbalance are not considered. Each slot can be in one of three states:
\begin{itemize}
\item{no packet's instances have been placed in a given slot, thus the slot is idle;}
\item{only 1 packet's instance has been placed in a given slot, thus the packet is correctly decoded;}
\item{more than 1 packet's instance has been placed in a given slot, thus resulting in interference of all involved packets.}
\end{itemize}

If at least one instance of a certain packet has been correctly received (see User 4), the contribution of the other instances of the same packet can be removed from the other slots. This process might allow to restore the content of packets in slots where intereference occurred (see User 2) and in an iterative manner other packets may be correctly decoded, up to a point in which no more packets can be restored or until the maximum number of iterations for the SIC process is reached.

Depending on the design choice, the packets that have not been decoded at the end of the SIC process are either discarded or a retransmission process is accomplished. In the latter case, a feedback is needed in order to inform users about the eventual failure of their transmission so that a certain retransmission policy can be applied. In the retransmission policy considered in this paper, each user retransmits its unsuccessful packet in one of the successive frames with probability $p_r$. In the remainder of this paper, we analyze the conditions that ensure stability of the overall transmission when using such a policy and we give a simple yet effective tool to design such a communication scenario in relation to the expected throughput and packet delay distribution gathered from the model.

\section{Stability Model}

Consider the aforementioned random access communication system. Each user can be in one of two states: Thinking (T) or Backlogged (B). Users in T state are idle users that generate a packet in a frame interval with probability $p_0$; if they do, no other packets are generated until successful transmission for that packet has been acknowledged. Users in B state are users that failed transmitting their packet, therefore they are waiting either to retransmit their unsuccessfully transmitted packet with probability $p_r$ at the beginning of each frame (thus geometrically distributed) or to receive a feedback about the outcome of their retransmission. Moreover, we assume that users are acknowledged about the success of their transmission at the end of the frame.

Let's define
\begin{itemize}
\item{$N_B^j$ : backlogged packets at the end of frame $j$}
\vspace{0.2cm}
\item{$G_B^j=\frac{N_B^{(j-1)}  p_r}{N_f}$ : expected channel load of frame $j$ due to users in B state}
\vspace{0.2cm}
\item{$G_T^j$ : expected channel load of frame $j$ due to users in T state}
\vspace{0.2cm}
\item{$G_{IN}^j=G_T^j+G_B^j$ : expected total channel load of frame $j$}
\vspace{0.2cm}
\item{$PLR^j(G_{IN}^j,N_f,d,I_{max})$ : expected average packet loss ratio of frame $j$, with dependence on the expected total channel load $G_{IN}$, the frame size $N_f$, the burst degree distribution $d$ and the maximum number of iterations for the SIC process $I_{max}$}
\vspace{0.2cm}
\item{$G_{OUT}^j=G_{IN}^j ( 1 - PLR(G_{IN}^j,N_f,d,I_{max}) )$ : part of load successfully transmitted in frame $j$, i.e. throughput.}
\end{itemize}
\vspace{0.2cm}

Our aim is to find the \textit{equilibrium contour} in the ($N_B$,$G_T$) plane \cite{STAB2}, defined as the locus of points for which at any frame, the expected channel load due to users in T state is equal to the expected throughput. Thus
\begin{equation}\label{gt}
 G_T=G_{OUT}=G_{IN} ( 1 - PLR(G_{IN},N_f,d,I_{max}) )
\end{equation}
where the frame number $j$ has been omitted, since in equilibrium state this condition must hold for any frame.
Moreover, being in an equilibrium point implies that also the expected number of backlogged users remains the same frame after frame. Therefore
\begin{equation}\label{nb1}
N_B=N_B (1-p_r) + G_{IN} N_f PLR(G_{IN},N_f,d,I_{max})
\end{equation}
from which
\begin{equation}\label{nb2}
N_B=\frac{G_{IN} PLR(G_{IN},N_f,d,I_{max}) N_f}{p_r}
\end{equation}
Equations \eqref{gt} and \eqref{nb2} completely describe the \textit{equilibrium contour}. In fact, once the configuration parameters $N_f$, $d$, $I_{max}$ of the system are set and the retransmission probability is chosen, the curves are completely described plotting $G_T$ and $N_B$ for different values of $G_{IN}$\footnote{Concerning the values used for the Packet Loss Ratio, it is known from the literature \cite{CRDSA1} \cite{IRSA1} that the relation between $PLR(G_{IN})$ and $G_{IN}$ can not be easily modelled in an analytical manner. For this reason PLR values used in this work are taken from simulations.} apart from the $M$ and $p_0$ considered.
\begin{figure}[t!]
\includegraphics [width=0.95 \columnwidth] {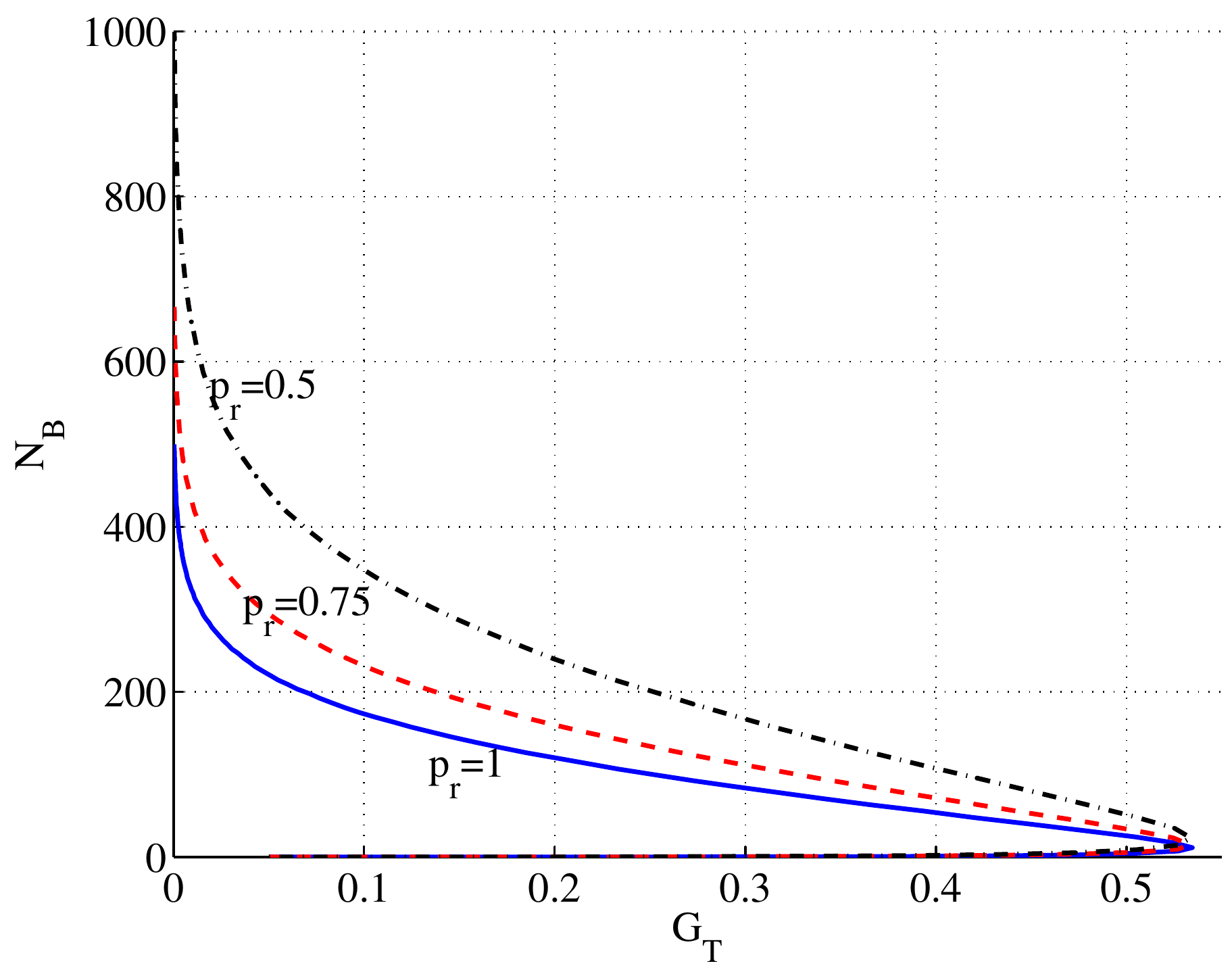}
\caption{\small{Equilibrium contours for CRDSA with $N_f=100 \ slots$, $I_{max}=20$}}
\label{Fig2}
\end{figure}
Figure~\ref{Fig2} displays examples of equilibrium contour for various $p_r$ values in the CRDSA case. As we can see, when the probability of retransmission on a given frame decreases, the equilibrium contour moves upwards so that the the maximum expected throughput is obtained for a bigger mean number of backlogged users $N_B$.

\subsection{Definition of Stability}

Let us study the conditions under which the described system is stable. In the followings we consider $M$ and $p_0$ to be constant (i.e. stationary input). Once $M$ and $p_0$ are defined, the expected channel input due to users in T state can be entirely described by the so called \textit{channel load line}, which represents the relation between $G_T$ and $N_B$ for a certain communication scenario. For the finite population case, the \textit{channel load line} can be defined as
\begin{equation}\label{LL1}
G_T=\frac{M-N_B}{N_f}p_0
\end{equation}
while for $M\rightarrow\infty$ the channel input can be described as a Poisson process with expected value $\lambda$ \cite{AbramsonALOHA2} so that $G_T=\lambda$ for any $N_B$, i.e. the expected channel input is constant and independent on the number of backlogged packets.

\begin{figure}[!]
\subfloat [Stable channel] {\label{stable} \includegraphics [ width =0.45 \columnwidth ]{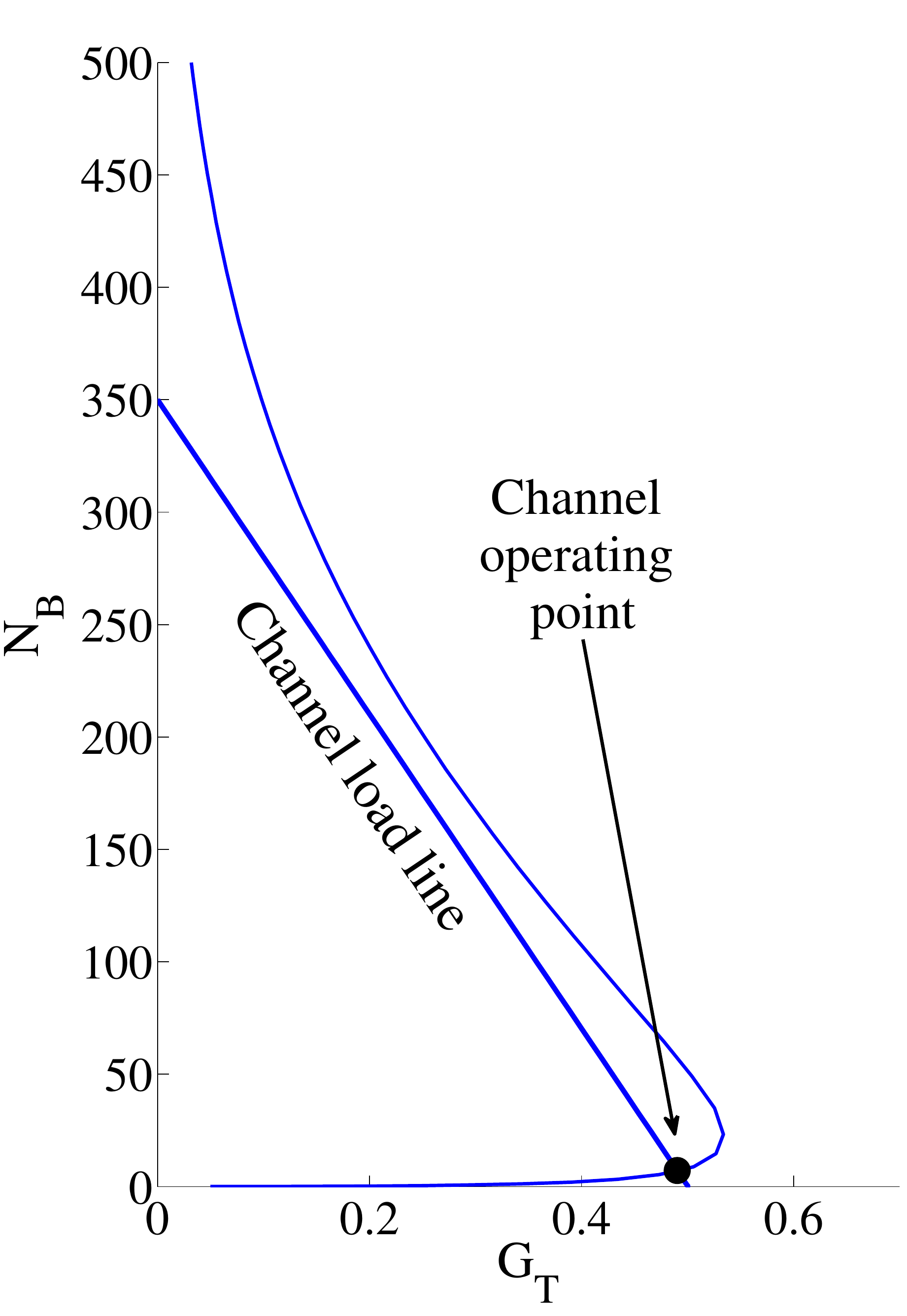}
\label{st}
} \qquad
\subfloat [Unstable channel (finite M)] {\label{unstableFin} \includegraphics [ width =0.45 \columnwidth ]{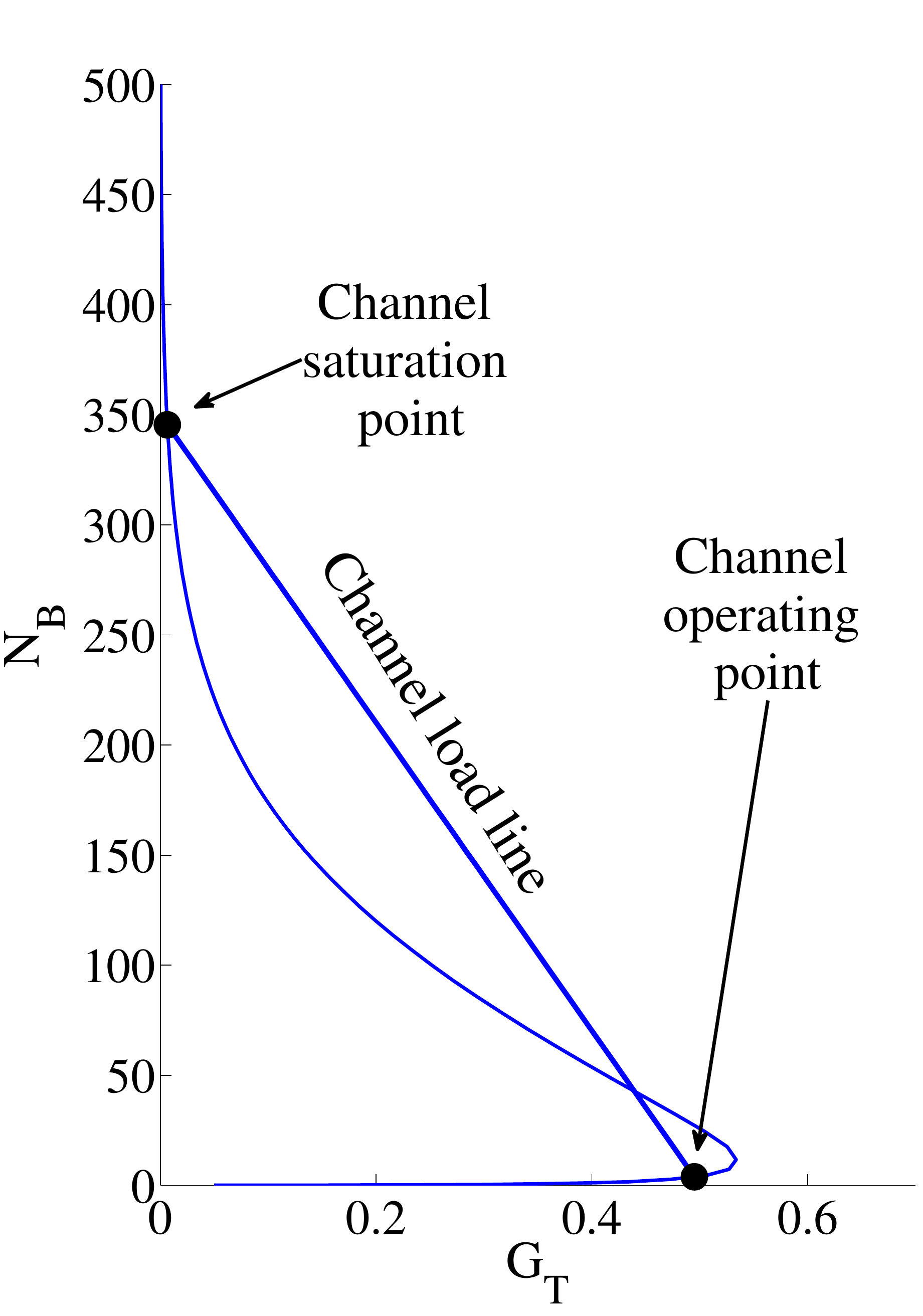}
\label{unstFin}
} \qquad
\\
\\

\begin{center}
\begin{tabular}{l l}

  $p_0=0.143$ & $p_0=0.143$ \\
\vspace{0.2cm}
  $p_r=0.5$ & $p_r=1$\\
\vspace{0.2cm}
  $M=350$ & $M=350$\\
\vspace{0.2cm}
  $(G_T^G,N_B^G)=(0.49,8.2) \hspace{1cm}$ & $(G_T^{S1},N_B^{S1})=(0.495,3.87)$\\
\vspace{0.2cm}
  & $(G_T^U,N_B^U)=(0.44,42.74)$\\
\vspace{0.2cm}
  & $(G_T^{S2},N_B^{S2})=(6\cdot 10^{-3},346.4)$\\
\hline

\end{tabular}
\end{center}

\subfloat [Unstable channel (infinite M)] {\label{unstableInf} \includegraphics [ width =0.45 \columnwidth ]{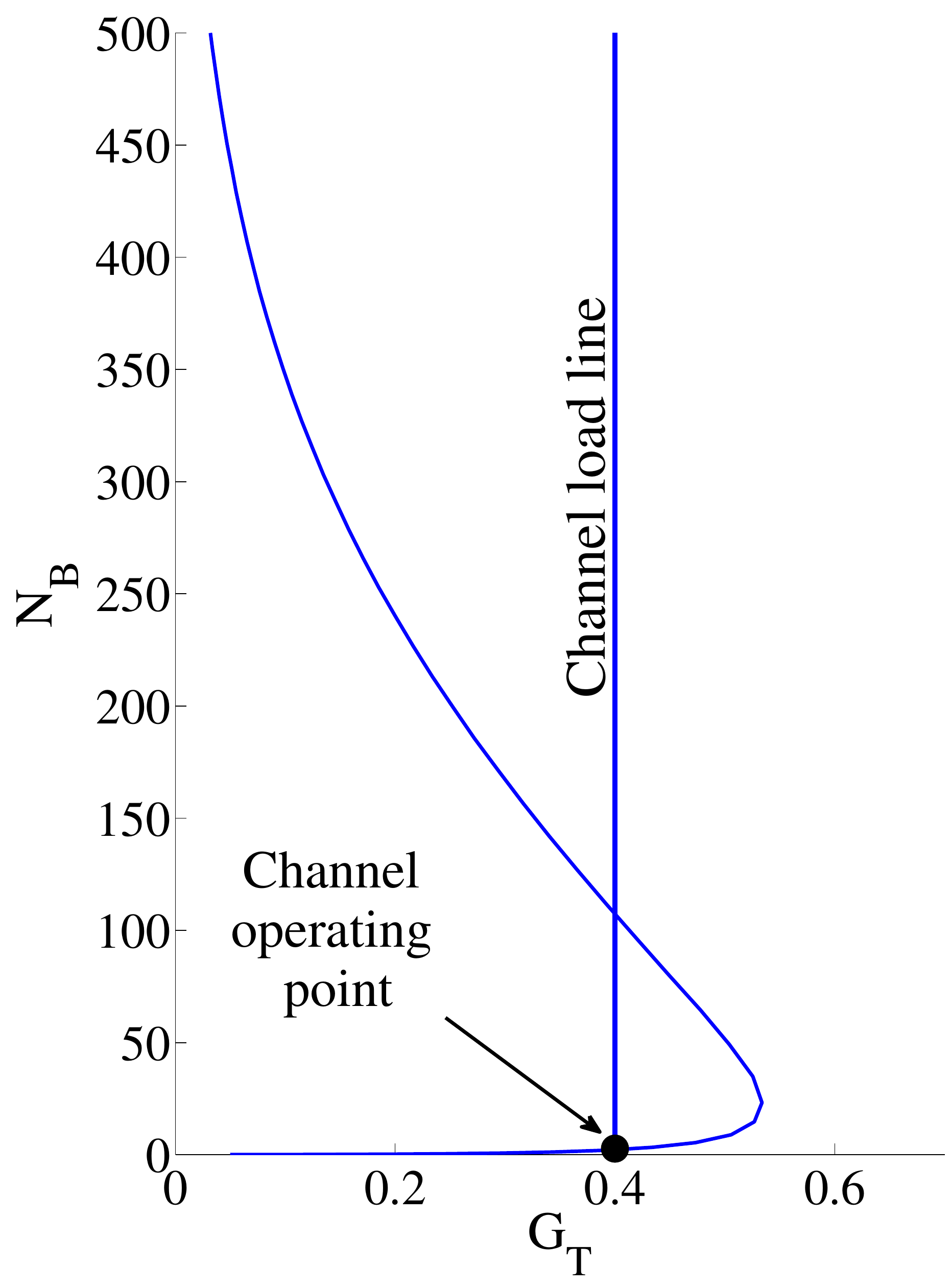}
\label{unstInf}
} \qquad
\subfloat [Overloaded channel] {\label{overL} \includegraphics [ width =0.45 \columnwidth ]{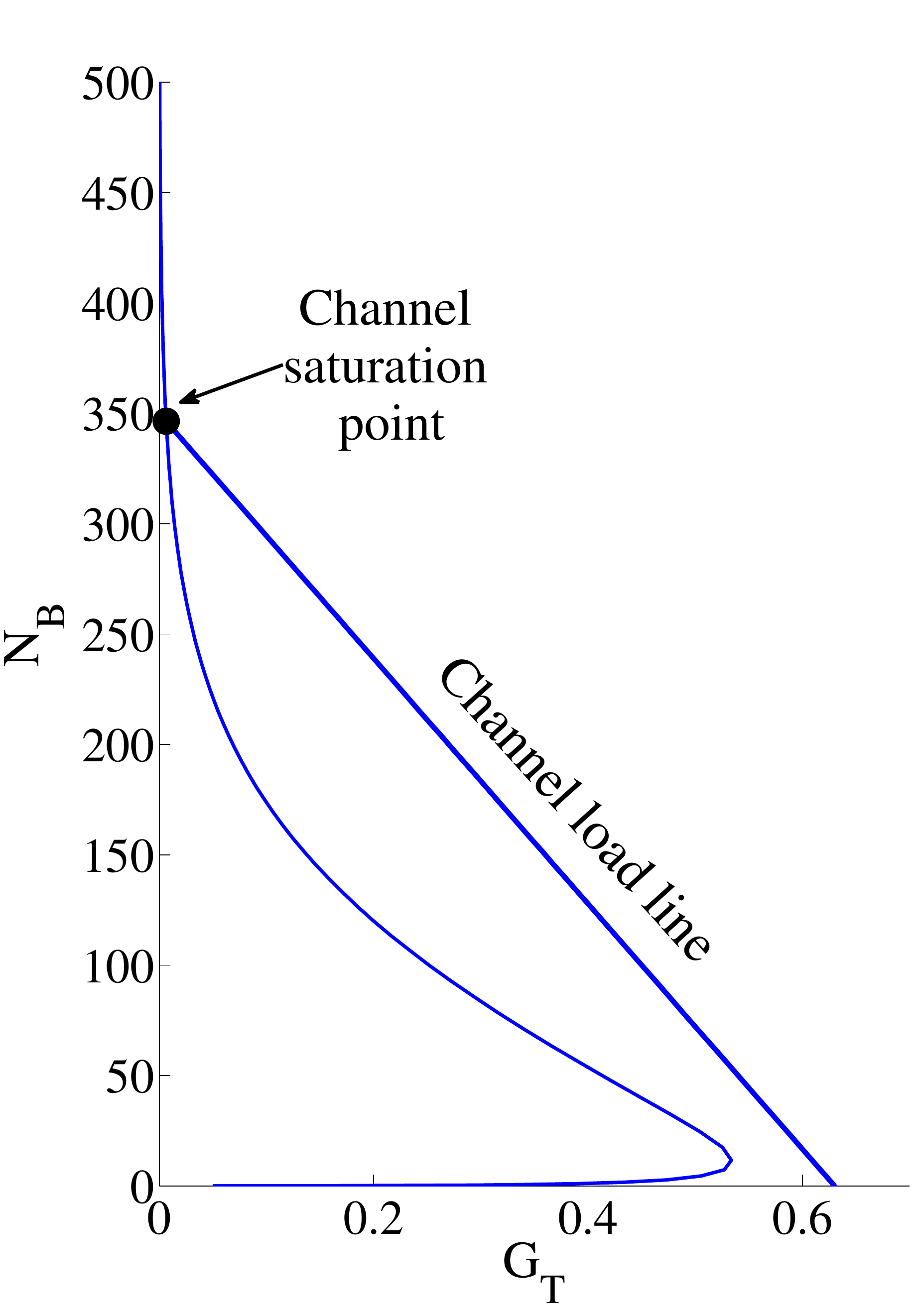}
\label{over}
} \qquad

\begin{center}
\begin{tabular}{l l}

  $\lambda=0.4$ & $p_0=0.18$ \\
\vspace{0.2cm}
  $p_r=0.5$ & $p_r=1$\\
\vspace{0.2cm}
  $M\rightarrow\infty$ & $M=350$\\
\vspace{0.2cm}
  $(G_T^{S1},N_B^{S1})=(0.4,1.7) \hspace{1.7cm}$ & $(G_T^S,N_B^S)=(6\cdot 10^{-3},346.4)$\\
\vspace{0.2cm}
  $(G_T^U,N_B^U)=(0.4,107.32)$ & \\
\vspace{0.2cm}
  $(G_T^{S2},N_B^{S2})=(0.4,\infty)$ & \\
\hline

\end{tabular}
\end{center}

\caption{\small{Examples of stable and unstable channels for CRDSA with $N_f=100$ and $I_{max}=20$. \textit{Stable equilibrium points} are marked with a black dot.}}
\label{All_channels}
\end{figure}
Consider Figure~\ref{All_channels}, representing various scenarios for CRDSA with $N_f=100\ slots$ and $I_{max}=20$. Equilibrium contours divide the ($N_B$,$G_T$) plane in two parts and each channel load line can have one or more intersections with the equilibrium contour. These intersections are referred to as equilibrium points. The rest of the points of the channel load line will belong to one of two sets: those on the left of the equilibrium contour represent points for which $G_{OUT}>G_T$, thus situations that yield to decrease of the backlogged population; those on the right represent points for which $G_{OUT}<G_T$, thus situations that yield to growth of the backlogged population.  

From the considerations above, we can gather that an intersection point where the channel load line enters the left part for increasing backlogged population corresponds to a \textit{stable equilibrium point}, since it acts as a sink. In particular, if the intersection is the only one, the point is a \textit{globally stable equilibrium point} (indicated as $G_T^G$,$N_B^G$), while if more than one intersection is present, it is a \textit{locally stable equilibrium point} (indicated as $G_T^S$,$N_B^S$). If an intersection point enters the right part for increasing backlogged population, it is said to be an \textit{unstable equilibrium point} (indicated as $G_T^U$,$N_B^U$) in the sense that as soon as a statistical variation from the equilibrium point occurs, the communication will diverge in one of the two directions of the channel load with equal probability (as claimed in \cite{KisslingStab} ).

Figure~\ref{st} shows a stable channel. The globally stable equilibrium point can be referred as \textit{channel operating point} in the sense that we expect the channel to operate around that point. With the word around we mean that due to statistical fluctuations, the actual $G_T$ and $N_B$ may differ from the expected value, however numerical results will show that averaging over the entire history of the transmission, values close to the expected ones are obtained. Figures~\ref{unstFin} and~\ref{unstInf} show unstable channels respectively for finite and infinite number of users. Analyzing this two figures for increasing number of backlogged packets, the first equilibrium point is a stable equilibrium point. Therefore the communication will tend to keep around it as for the stable equilibrium point in Figure~\ref{st}, and we can refer to it once again as \textit{channel operating point}. However, this is not the only point of equilibrium since more intersections are present. Therefore, due to the abovementioned statistical fluctuations, the number of backlogged users could pass the second intersection and return to the right part of the plane, causing an unbounded increase of the expected number of backlogged users in the case of infinite $M$ (Figure~\ref{unstInf}) or an increase till a new intersection point is reached in the case of finite $M$ (Figure~\ref{unstFin}). In the latter case, this third intersection point is another stable equilibrium point known as \textit{channel saturation point}, so called because it is a condition in which almost any user is in B state and $G_{OUT}$ approaches zero. In the former case of infinite $M$, $N_B$ will increase indefinitely and we can say that a channel saturation point is present for $N_B\rightarrow\infty$. Notice that if $M$ is finite, a stable channel can always be achieved using a sufficiently small value of $p_r$ (the channel in Figure~\ref{st} has the same parameters as the one in Figure~\ref{unstFin} except for $p_r$ that is $0.5$ instead of $1$). However, when $p_r$ gets smaller, the corresponding average packet delay could get larger as we will show in Section V. Therefore a tradeoff between stability of the channel and average packet delay need to be faced in the design phase. Finally Figure~\ref{overL} shows the case of an overloaded channel. In this case there is only one equilibrium point corresponding to the channel saturation point. As for unstable channels with finite M, this channel can be rendered stable decreasing $p_r$. Now, from the equilibrium contour we know that the point for the maximum expected throughput is ($G_T^{max}$,$N_B^{max}$), therefore the communication channel can be designed using Equation \eqref{LL1}:

\begin{equation}\label{LLmax}
G_T^{max}=\frac{M-N_B^{max}}{N_f}p_0
\end{equation}

given that $(M / N_f)\cdot p_0 \geq G_{max}$ must hold since the slope (known in the literature as $m$) of the channel load line must be a negative value. As an example, if $p_0$ is a fixed and not modifiable design constraint, the number of users M that ensures maximum throughput can be calculated from \eqref{LLmax}. However the stability and the average packet delay need to be computed in order to verify they fulfill the design constraints. The former can be verified as illustrated above, the latter can be computed as shown in the next section.

\section{Packet Delay Model}
\begin{figure}[h!]
\centering
\includegraphics [width=0.95 \columnwidth] {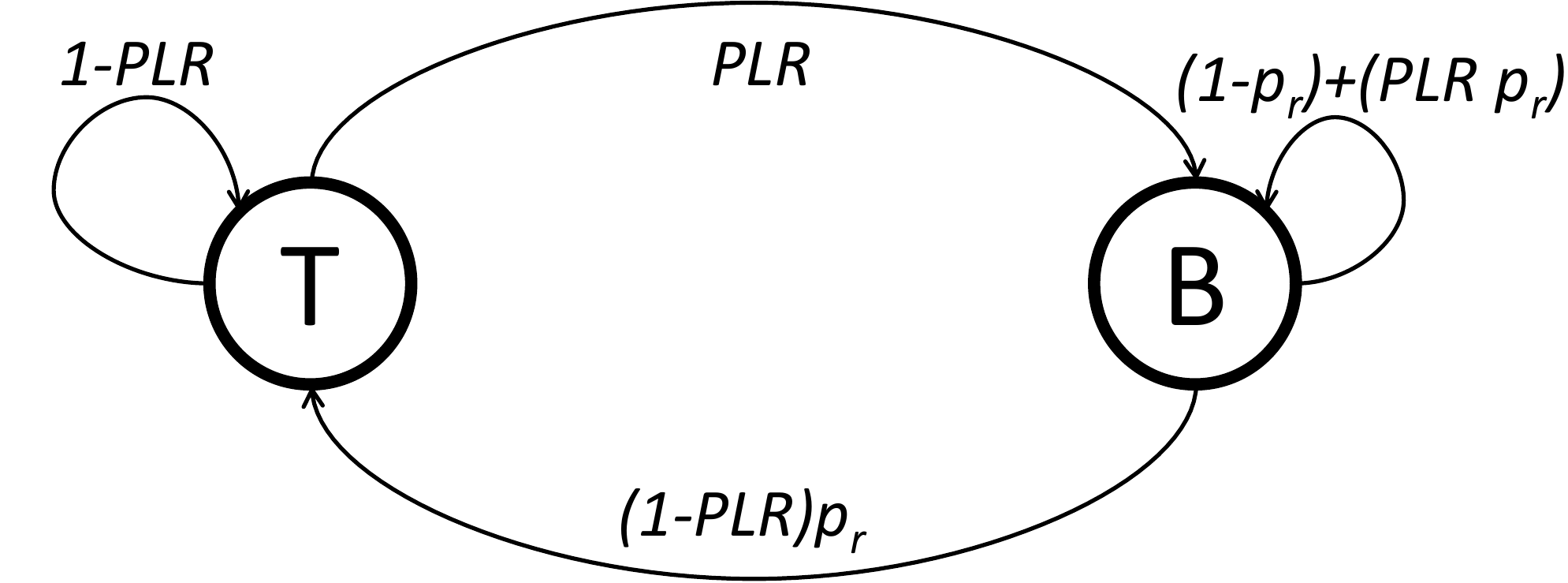}
\caption{\small{Markov Chain for the Packet Delay analysis}}
\label{FSM}
\end{figure}

Assuming that a channel is operating at its \textit{channel operating point}, we would like to know what is the expected distribution and mean delay associated to successfully transmitted packets. This can be entirely described using a discrete-time Markov chain with two states (Figure~\ref{FSM}). The two states represent the state of a generic user that has a packet to transmit at the end of each frame. Therefore a frame duration is our discrete time unit for this Markov chain. The edges emanating from the states represent the state transitions occurring to users. These transitions depend on $p_r$ and $PLR(G_{IN}^O,N_f,d,I_{max})$. If a packet transmitting for the first time receives a positive acknowledgement (ACK) the user stays in T state while in case of negative acknowledgement (NACK) the user switches to B state until successful retransmission has been achieved and acknowledged. Therefore, here the packet delay $D_{pkt}$ is considered as the number of frames that elapse from the beginning of the frame in which the packet was transmitted for the first time, till the end of the one in which the packet was correctly received.

According to the definition given above, the delay distribution is entirely described by

\begin{equation}\label{FSMeq}
Pr\{D_{pkt}=n\} =
\begin{cases}
1-PLR\text{\,,\ \ \ \ \ \ \ \ \ \ \ \ \ \ \ \ \ \ \ \ \ for  } n=1 \\
\\
PLR\ [p_r\ (1-PLR)] \cdot \\
\cdot [1-p_r+PLR\ p_r]^{n-2}     \text{\ ,\ \ \ \ for  } n>1
\end{cases}
\end{equation}

while the average expected delay is

\begin{equation}\label{FSMAvEq}
Av[D_{pkt}]=\sum_{n=1}^{\infty}  n \cdot Pr\{D_{pkt}=n\}
\end{equation}

\begin{figure}[h!]
\centering
\includegraphics [width=1 \columnwidth] {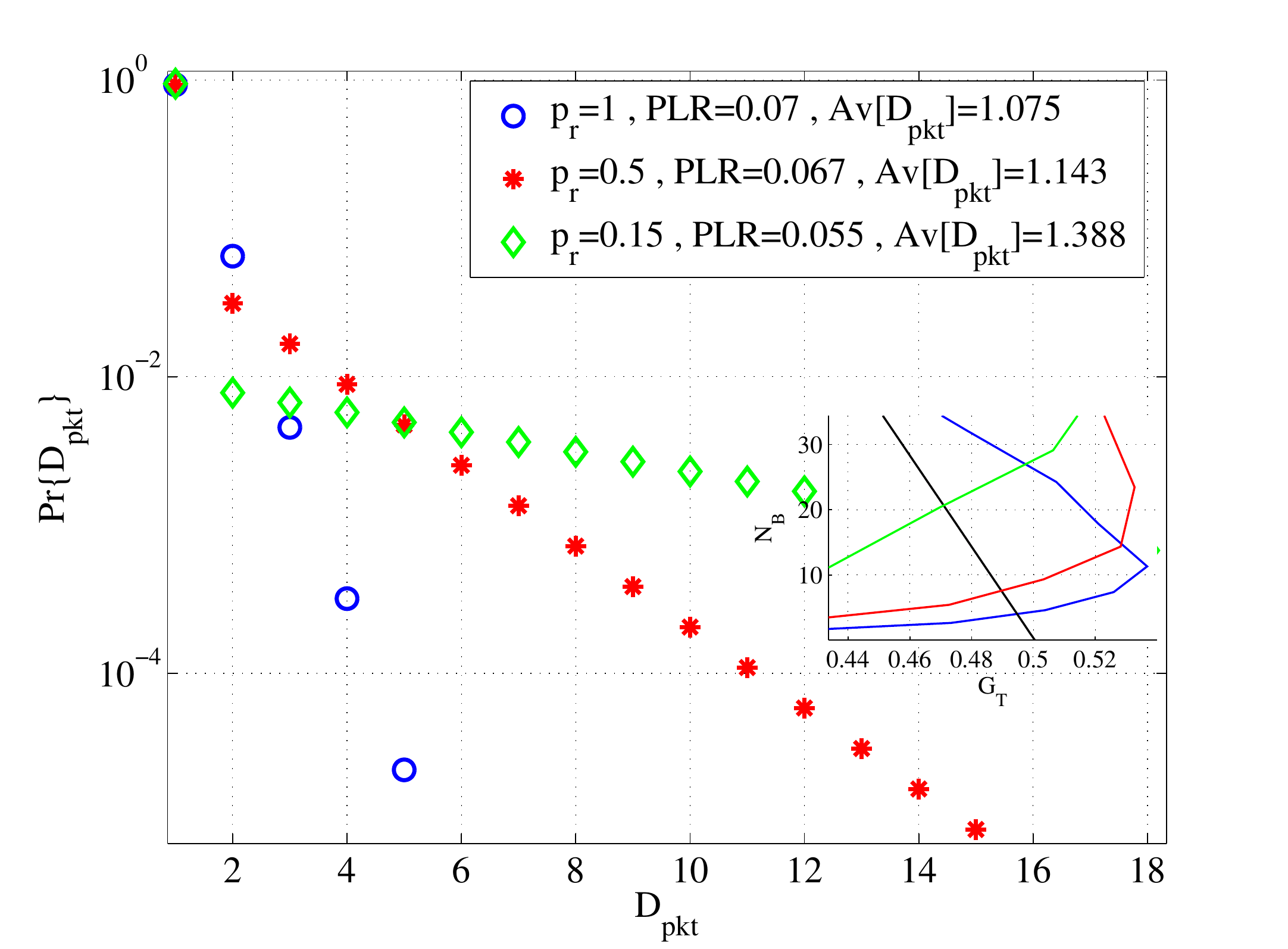}
\caption{\small{Delay distribution for CRDSA with $N_f=100 \ slots$,\\ $I_{max}=20$, $M=350$ and $p_0=0.143$}}
\label{del_dist}
\end{figure}

Figure~\ref{del_dist} shows some examples of delay distribution. In this particular case, when $p_r$ gets smaller the distribution spreads over bigger values of delay and even though the probability that a packet is successfully transmitted at the first attempt increases, the expected average packet delay gets higher. This can be also explained noting that in this example, when decreasing $p_r$ the throughput decreases while the number of backlogged packets increases so that a packet will require more time to be successfully transmitted. Let us call $G_{OUT}^{max}$ the maximum throughput achievable. If $M$ and $p_0$ are such that $(M\cdot p_0)/N_f\leq G_{OUT}^{max}$, the average packet delay at the channel operating point will always increase when decreasing $p_r$, since the equilibrium contour moves upwards.

\section{Numerical Results}
In this section, the results of simulations are shown in order to validate the stability model described above. The simulator has been built according to the system description given in Section II, therefore perfect interference cancellation and channel estimation have been assumed. Moreover, neither the possibility of FEC nor power unbalance have been considered for our simulations.

\begin{figure}[h!]
\includegraphics [width=0.95 \columnwidth] {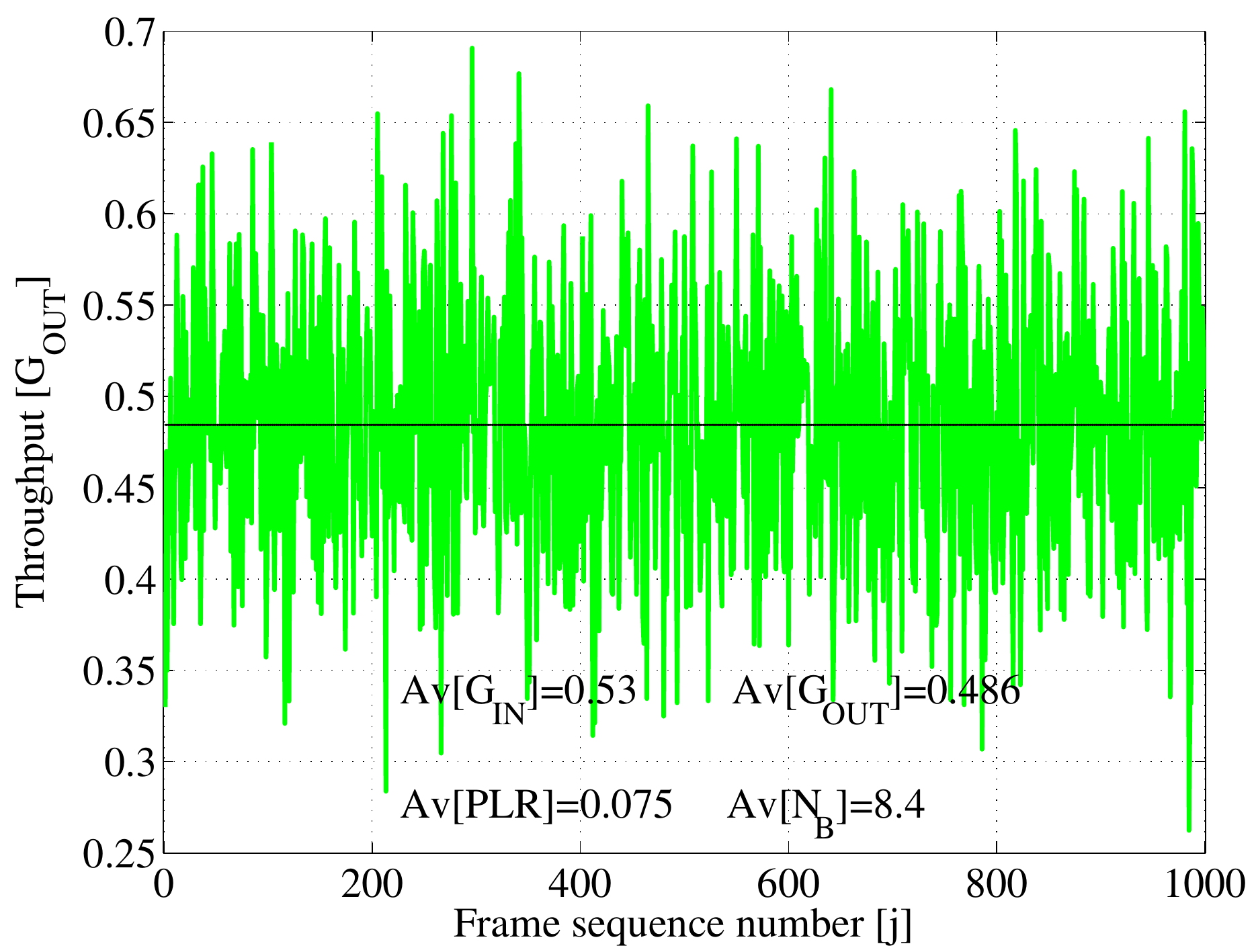}
\caption{\small{Simulated throughput for CRDSA with $N_f=100 \ slots$, $I_{max}=20$, $p_0=0.143$, $p_r=0.5$, $M=350$}}
\label{st_sim}
\end{figure}

Figure~\ref{st_sim} shows the result of simulations for a communication scenario with the same parameters as the example of stable channel described in Figure~\ref{st}. It can be seen that due to the aforementioned statistical variations, the throughput oscillates around a certain value. This value is the equilibrium point. In fact the horizontal line represents the value $Av[G_{OUT}]=0.486$, that is the value obtained averaging the throughput over the entire simulation and is, as expected, a value really close to the one claimed in Figure~\ref{st}. Moreover, also the related average packet delay has been found to agree with the analytical results.\footnote{In the simulations only the average packet delay has been computed. In fact, regarding the delay distribution, a more complicated simulator is required in order to trace the identity of each packet so that each individual packet delay is known. Even though this can be accomplished, we have not built such a simulator yet.}

\begin{figure}[h!]
\includegraphics [width=0.95 \columnwidth] {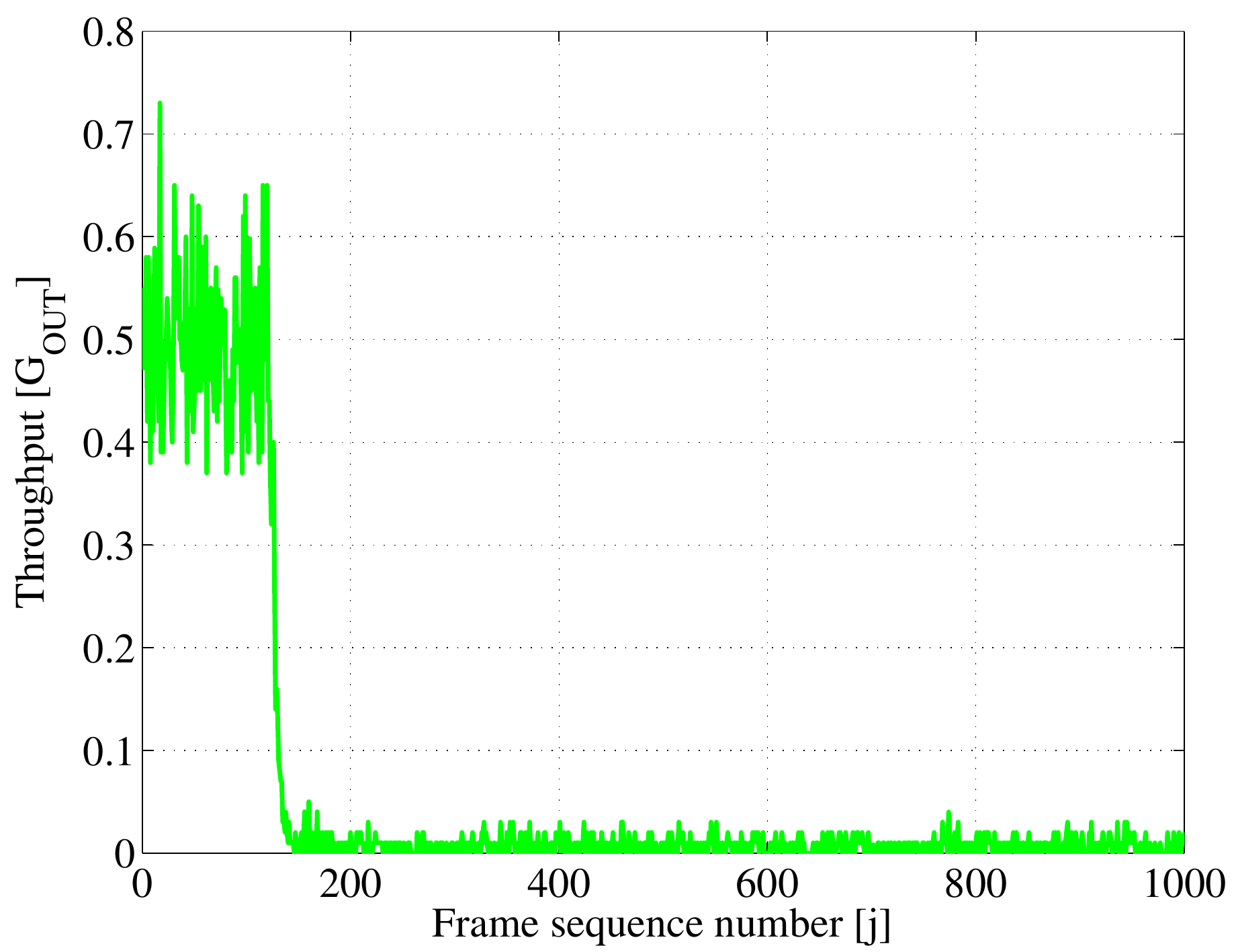}
\caption{\small{Simulated throughput for CRDSA with $N_f=100 \ slots$, $I_{max}=20$, $p_0=0.143$, $p_r=1$, $M=350$}}
\label{unstFin_sim}
\end{figure}

Figure~\ref{unstFin_sim} illustrates an outcome of simulations for the communication scenario of the unstable channel described in Figure~\ref{unstFin}. In this case, the initial behaviour of the channel is such that the throughput oscillates around the channel operating point. However, differently from the previous example, this is not a globally stable equilibrium point. Therefore statistical fluctuations will sooner or later cause divergence from the channel operating point and the subsequent saturation of the channel with an average throughput that approaches zero. The time it takes for the channel to diverge from the channel operating point varies from simulation to simulation. In literature \cite{STAB2} the expected time of divergence for Slotted Aloha is known as First Exit Time (FET). In our case this value is not easily computable since a Markov chain describing the communication and considering all the transition probabilities needs to be computed. This is computationally costly (as already claimed in \cite{KisslingStab}) and we have not carried on methods and formulas for its calculation yet.

\begin{figure}[h!]
\includegraphics [width=0.95 \columnwidth] {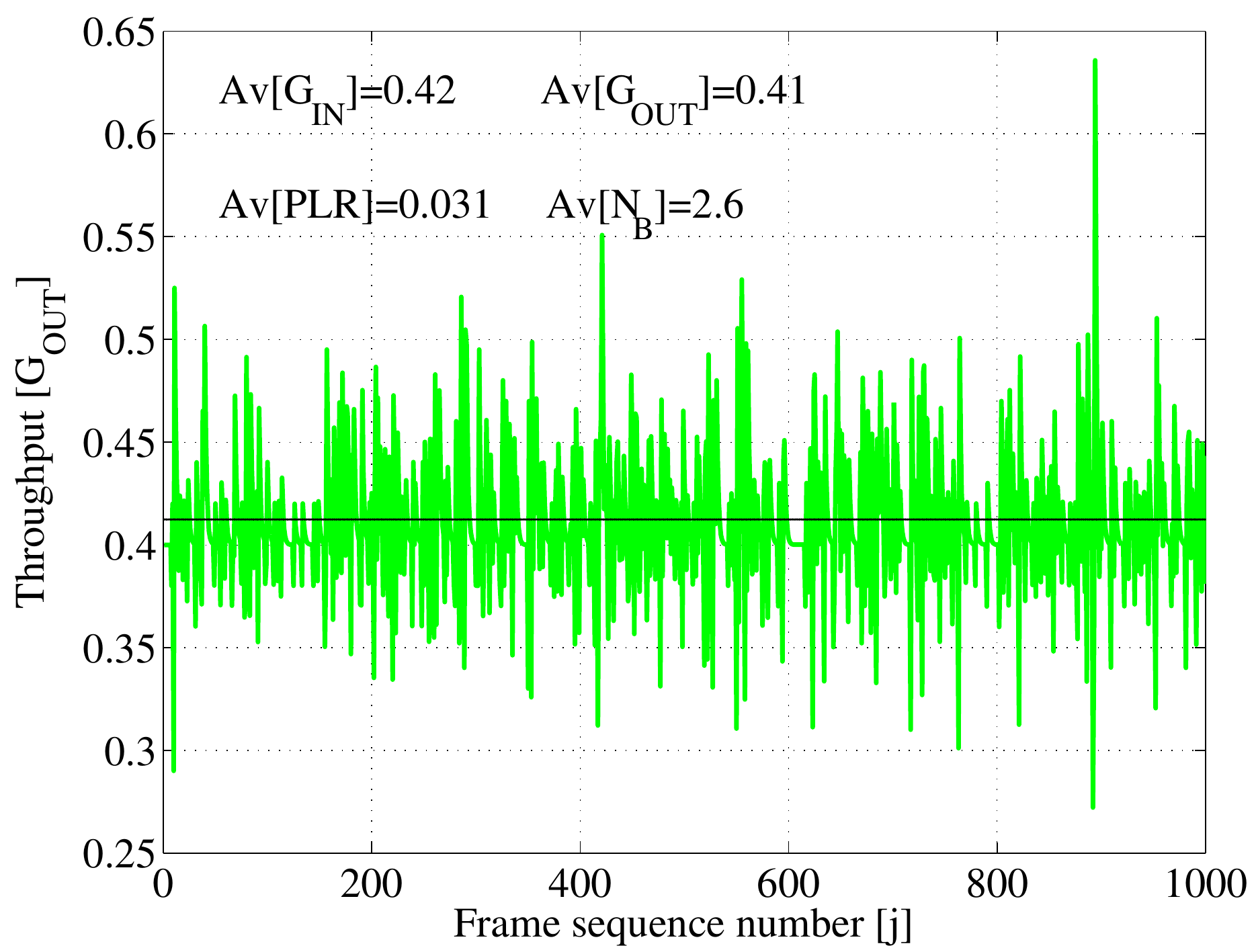}
\caption{\small{Simulated throughput for CRDSA with $N_f=100 \ slots$, $I_{max}=20$, $\lambda=0.4$, $p_r=0.5$, $M\rightarrow\infty$ in the case before divergence from the operating point occurred}}
\label{unstInf_sim-ok}
\end{figure}

\begin{figure}[h!]
\includegraphics [width=0.95 \columnwidth] {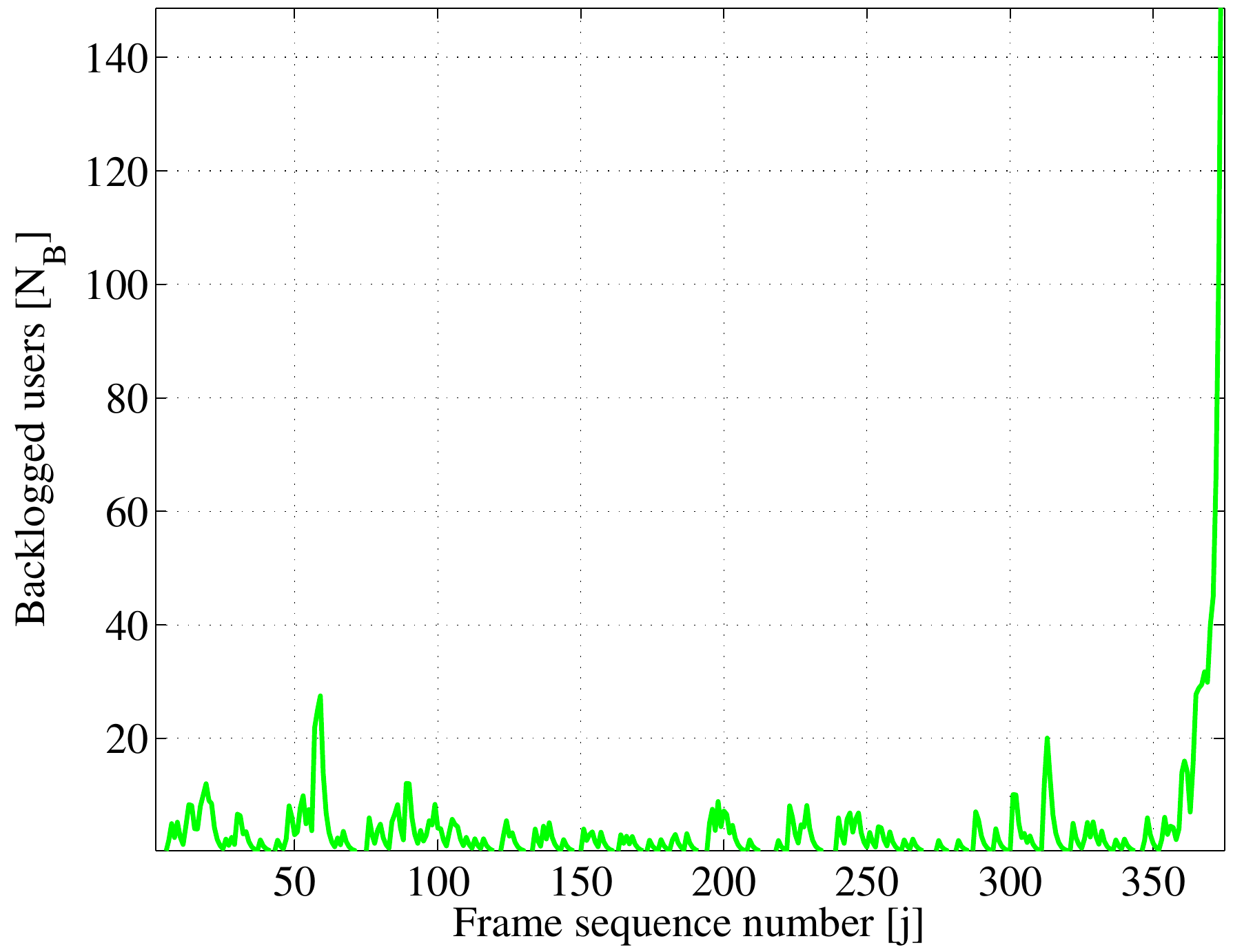}
\caption{\small{Number of backlogged users for CRDSA with $N_f=100 \ slots$, $I_{max}=20$, $\lambda=0.4$, $p_r=0.5$, $M\rightarrow\infty$ with divergence from the operating point after a certain time}}
\label{unstInf_sim-BL_ko}
\end{figure}

Finally, Figure~\ref{unstInf_sim-ok} illustrates an outcome of simulations for the case illustrated in Figure~\ref{unstInf}, that is the case of infinite population. In particular, this outcome shows an occurrence in which divergence from the channel operating point has not occurred yet. As we can see, as long as the communication takes place around the channel operating point, the same throughput and delay considerations as for the stable channel are valid. This example highlights that depending on the communication parameters, even though the channel is unstable, the FET could be so big that instability might be acceptable depending on application. However, if this is not the case, the communication will soon exit the stability region and  the number of backlogged users will grow fast and indefinitely as shown in Figure~\ref{unstInf_sim-BL_ko}.

\section{Conclusions}
In this paper a model for CRDSA based on equilibrium contour and extandable also to CRDSA++ and IRSA has been proposed. This model allows to study and predict the stability and to calculate the expected throughput as well as the average and expected distribution delay at the channel operating point, thus allowing a design that based on design constraints considers the tradeoff between stability, throughput and delay in order to provide an efficient and stable communication. As also suggested by reviewers (that we gratefully acknowledge for their suggestions and comments), future work will regard the analysis of a broader range of cases, the use of a different number of replicas than 2 in order to better understand this dependency and the study of such a communication scenario when introducing FEC codes as well as the possibility of packet power unbalance.


\begin{thebibliography}{1}

\bibitem{AbramsonALOHA1}
N.~Abramson, "The aloha system: Another alternative for computer communications", in \emph{Proceedings of the 1970 Fall Joint Comput. Conf., AFIPS Conf.}, vol.37, Montvale, N.J., 1970, pp.281-285.

\bibitem{AbramsonALOHA2}
N.~Abramson, "The throughput of packet broadcasting channels", \emph{IEEE Trans.Comm.}, vol.25, pp.117-128, Jan. 1977.

\bibitem{RobertsALOHA}
L.G.~Roberts, "ALOHA packet systems with and without slots and capture", ARPANET System Note 8 (NIC11290), June 1972.

\bibitem{DiversityALOHA}
G.L.~Choudhury and S.~S.~Rappaport, "Diversity ALOHA - A random access scheme for satellite communications", \emph{IEEE Trans.Comm.}, vol.31, pp.450-457, Mar. 1983.

\bibitem{STAB1}
Carleial, A.; Hellman, M.; , "Bistable Behavior of ALOHA-Type Systems," Communications, IEEE Transactions on , vol.23, no.4, pp. 401- 410, Apr 1975

\bibitem{STAB2}
Kleinrock, L.; Lam, S.; , "Packet Switching in a Multiaccess Broadcast Channel: Performance Evaluation," Communications, IEEE Transactions on , vol.23, no.4, pp. 410- 423, Apr 1975

\bibitem{STABcontrol}
Lam, S.; Kleinrock, L.; , "Packet Switching in a Multiaccess Broadcast Channel: Dynamic Control Procedures," Communications, IEEE Transactions on , vol.23, no.9, pp. 891- 904, Sep 1975

\bibitem{CRDSA1}
Casini, E.; De Gaudenzi, R.; Herrero, Od.R.; , "Contention Resolution Diversity Slotted ALOHA (CRDSA): An Enhanced Random Access Schemefor Satellite Access Packet Networks," Wireless Communications, IEEE Transactions on , vol.6, no.4, pp.1408-1419, April 2007

\bibitem{CRDSA2}
De Gaudenzi, R.; del Rio Herrero, O.; , "Advances in Random Access protocols for satellite networks," Satellite and Space Communications, 2009. IWSSC 2009. International Workshop on , vol., no., pp.331-336, 9-11 Sept. 2009

\bibitem{IRSA1}
Liva, G.; , "Graph-Based Analysis and Optimization of Contention Resolution Diversity Slotted ALOHA," Communications, IEEE Transactions on , vol.59, no.2, pp.477-487, February 2011

\bibitem{KisslingStab}
Kissling, C.; , "On the Stability of Contention Resolution Diversity Slotted ALOHA (CRDSA)," Global Telecommunications Conference (GLOBECOM 2011), 2011 IEEE , vol., no., pp.1-6, 5-9 Dec. 2011

\end{thebibliography}
\end{document}